\documentstyle[preprint,psfig,aps]{revtex}

\def\be{\begin{equation}}
\def\ee{\end{equation}}
\begin{document}


\title{Adiabatic reduction near a bifurcation in  stochastically modulated 
systems}

\author{Fran\c{c}ois Drolet$^{1}$ and Jorge Vi\~nals$^{1,2}$}
\address{$^{1}$ Supercomputer Computations Research Institute, Florida State
University, Tallahassee, Florida 32306-4130. $^{2}$ Department of Chemical
Engineering, FAMU-FSU College of Engineering, Tallahassee, Florida 32310-6046}

\date{\today}
   
\maketitle

\begin{abstract}

We re-examine the procedure of adiabatic elimination of fast relaxing
variables near a bifurcation point when some of the parameters of the system 
are stochastically modulated.
Approximate stationary solutions of the Fokker-Planck equation 
are obtained near threshold for the pitchfork and transcritical bifurcations.
Stochastic resonance between fast variables and random modulation
may shift the effective bifurcation point by an amount proportional to
the intensity of the fluctuations. We also find that
fluctuations of the fast variables above
threshold are not always Gaussian and centered around the (deterministic)
center manifold as was previously believed. Numerical solutions obtained for 
a few illustrative examples support these conclusions.
\end{abstract}


\section{Introduction}

A system is said to undergo a bifurcation when its long time behavior changes
qualitatively as some control parameter is continuously varied. Examples
include the saddle-node, transcritical and pitchfork bifurcations, which
involve a transition between two fixed point solutions, and the Hopf
bifurcation that involves a transition between a fixed point solution and a
limit cycle. Near the bifurcation point only a small number of so-called
slow variables are required to determine the evolution of the system over
a long time scale. The remaining degrees of freedom (the so-called fast
variables) adjust very rapidly to the instantaneous values of the slow
variables, and can be adiabatically eliminated. The qualitative features of 
the evolution of the system near the bifurcation point are thus obtained by 
constraining the original governing
equations to a surface in phase space known as the center manifold. The
resulting equations valid on the manifold are the normal form
equations \cite{re:guckenheimer83}. The purpose of this article is
to re-examine the analogous reduction procedure when one or more of the 
system's parameters include a random component 
\cite{re:stratonovich67,re:haken77,re:vandenbroeck82,%
re:knobloch83,re:elphick87,re:drolet97}.

We focus mainly on the case in which the externally set control parameter
includes a small random component which we model as a stochastic process in
time. In this case, the bifurcation point remains sharp,
although its location may depend on the intensity of the fluctuations. 
Although there is arguably little conceptual difference 
between deterministic variables that relax quickly in the vicinity of the 
bifurcation point, and a stochastic process of short correlation time 
(say of the same order or smaller than inverse relaxation rates of the fast 
variables), we show below that stochastic resonance between the two can 
affect the evolution on the slow time scale.

The essential aspects of the adiabatic reduction procedure in the 
stochastic case can be illustrated in the simple case of a second order
system. Let $A$ be the amplitude of a bifurcating mode, and $B$ the 
amplitude of a second mode that is itself linearly stable near onset. 
A reduced control parameter $\alpha$ is defined such that the trivial
state $A=B=0$ is stable if $\alpha \leq 0$, and unstable otherwise. 
Fluctuations in $\alpha$ are included through a stochastic process 
$\xi(t)$, which we assume Gaussian, white and of small intensity $\kappa$. 
The evolution of the system is now stochastic and is described by the
joint probability density ${\cal P}(A,B;t)$ at time $t$. The reduction
procedure starts by decomposing the joint density as
\begin{equation}
\label{eq:bayes}
{\cal P}(A,B;t) = p(B|A;t)P(A;t),
\end{equation}
where $p(B|A;t)$ is the conditional probability density. Close to 
threshold, the stochastic processes $A$ and $B$ are small (their
intensity scales with some power of $\kappa$) in such a way that
characteristic values of $B/A \sim \kappa^{a} \ll 1, ~~ a>0$. As will be shown
in more detail below, this assumption also implies that the two processes
evolve over different characteristic temporal scales, fact that is
reminiscent of the separation of time scales present in the 
deterministic limit. As a consequence, the probability densities 
$P(A;t)$ and $p(B|A;t)$ can be separately obtained at different orders in 
$\kappa$. The stationary density $P(A)$ is then used to locate the 
effective threshold point in the stochastic case. Below threshold, $P(A)$ is a
delta function at $A=0$, whereas above threshold there exists another
normalizable solution that has some non vanishing moments. 

Van den Broeck et al. \cite{re:vandenbroeck82} introduced this reduction
procedure to study the effect of additive noise on a pitchfork bifurcation.
They derived an approximate
expression for the stationary probability density near but {\em below}
threshold.  They  showed that in the weak noise limit, the critical 
variable exhibits amplified
non-Gaussian fluctuations and that the properties of the fast variable 
depend on the nonlinearity of the system under study. 
Their analysis, however, is difficult to extend to the region above 
threshold. We find that additive noise eliminates 
the separation in scales between the slow and fast variables, and that,
as a consequence, the probability densities for $A$ and $B$
are in general quite broad. Hence the assumption that $A/B \ll 1$ breaks 
down over significant portions of any particular trajectory, and the reduction
procedure discussed is not reliable.

In view of this limitation, the analysis presented here is restricted
to equations involving multiplicative noise only. In this case,
the separation in scales between the fast and slow variables is 
preserved well above onset. Knobloch and Wiesenfeld 
\cite{re:knobloch83} had already addressed the
adiabatic elimination procedure in the multiplicative case by 
introducing one additional assumption: that fast variables
are gaussianly distributed around the underlying deterministic
center manifold. Our analysis extends theirs in that such an assumption
is not necessary. In fact, we show that the fast variable does not always 
fluctuate around the manifold.

We derive approximate expressions for the stationary
probability densities $p(B|A)$ and $P(A)$ valid near threshold for 
the pitchfork and
transcritical bifurcations. In both cases, the marginal density
$P(A)$ has to satisfy a normalizability condition that is used to 
determine the location of onset $\alpha_c$. In those cases in which
$\alpha_{c} \ne 0$, stochastic resonance between 
the fast variable $B$ and the stochastic process $\xi(t)$ is responsible 
for the shift away from the deterministic threshold. This result
generalizes earlier analyses of the normal form equation corresponding to
a pitchfork bifurcation with a fluctuating control parameter 
\cite{re:schenzle79,re:graham82b}, in which coupling to fast variables was not 
considered. In agreement with our results below, the absence of such coupling 
leads to $\alpha_{c} = 0$ for any intensity of the fluctuating control
parameter. 

The case of a pitchfork bifurcation with multiplicative noise
is considered in Section \ref{sec:vdp}. For simplicity, the method is
applied to the well-known Van der Pol-Duffing equation. In that example,
the bifurcation point is shifted to $\alpha_c > 0 $, while the fast variable
$B$ exhibits Gaussian fluctuations around $B = 0$. Our result for 
$\alpha_{c}$ agrees 
with earlier work by L\"ucke \cite{re:lucke89}, but
disagrees with the work of Knobloch and Wiesenfeld \cite{re:knobloch83}
and of Seshadri et al. \cite{re:seshadri81}.
Section \ref{sec:transc} considers the general case of a transcritical
bifurcation. In this case, the fast variable can exhibit non Gaussian
fluctuations and, in general, the mean of the distribution
does not lie on the underlying deterministic center manifold.

\section{Pitchfork bifurcation with multiplicative noise : the Van Der 
 Pol-Duffing Equation}
\label{sec:vdp}

In order to illustrate the reduction procedure in a model system that 
bifurcates supercritically, we consider the 
non-linear oscillator 
\begin{eqnarray}
\label{eq:vdp4}
    \frac{d}{dt} \left[ \begin{array}{c} x \\ \dot{x}
        \end{array}  \right]  & = & \left[ \begin{array}{cc}
         0 &  1  \\
        \alpha  & - \beta \end{array} \right]
        \left[ \begin{array}{c} x \\ \dot{x} \end{array} \right] +
        \left[ \begin{array}{c}
           0  \\ - a x^3 - b x^2 \dot{x}
    \end{array} \right] + \left[ \begin{array}{cc} 0 & 0 \\
    1 & 0 \end{array} \right] \left[ \begin{array}{c} x \\ \dot{x}
     \end{array} \right] \xi(t),
\end{eqnarray}
also known as Van der Pol-Duffing oscillator \cite{re:wiesenfeld82}. 
The positive constants $\beta, a$ and $b$ are of ${\cal O}(1)$.
At $\alpha=0$, Eq. (\ref{eq:vdp4}) exhibits a supercritical pitchfork
bifurcation between the two fixed point solutions $x=0$ (stable for
$\alpha < 0$) and
$x=\pm \sqrt{\alpha/a}$ (stable for $\alpha > 0$). The last term in the 
right-hand side originates
from a random component in the control parameter $\alpha$. We limit our 
analysis to Gaussian, white noise satisfying $\langle \xi(t) \rangle = 0$ 
and $\langle \xi(t) \xi(t') \rangle = 2 \kappa \delta (t-t')$, where
$\langle ~ \rangle$ denotes an ensemble average and $\kappa$ is the 
intensity of the noise.  
Motivated by the 
known center manifold reduction in the deterministic limit of $\kappa =0$,
we perform the following linear change 
of variables $A=x + \dot{x} /\beta$ and $B = - \dot{x} / \beta$ 
to yield \cite{re:knobloch83},
\begin{eqnarray}
\label{eq:vdp5}
\nonumber
    \frac{d}{dt} \left[ \begin{array}{c} A \\  B
        \end{array}  \right]  & = & \left[ \begin{array}{cc}
         \alpha / \beta &  \alpha / \beta  \\
        - \alpha/ \beta  & - (\beta + \alpha/\beta) \end{array} \right]
        \left[ \begin{array}{c} A \\ B \end{array} \right] +
        \left[ \begin{array}{c}
           -c A^3  + d A^2 B + e A B^2 + f B^3  \\
           +c A^3  - d A^2 B - e A B^2 - f B^3 \end{array} \right]
         \\  & &
      + \left[ \begin{array}{cc} 1 & 1 \\ -1 & -1 \end{array} \right]
        \left[ \begin{array}{c} A \\ B \end{array} \right]
        \frac{\xi(t)}{\beta},
\end{eqnarray}
with $c=a/\beta, d=b-3a/\beta, e=2b - 3a/\beta$ and $f=b - a/\beta$.
The linear matrix $L \equiv \left[ \begin{array}{cc}
 \alpha/\beta & \alpha/ \beta  \\  -\alpha/\beta  & - (\beta + \alpha/\beta)
 \end{array} \right]$ has a zero eigenvalue at the 
deterministic bifurcation point of $\alpha = 0$,
with a second eigenvalue of ${\cal O}(1)$. In the absence
of noise, the variable $B$ thus varies over a time scale which is much
faster than the time scale of $A$. One then 
introduces the scalings $\alpha \sim {\cal O} (\epsilon^{2})$, $T =
\epsilon^{2} t, A \sim {\cal O}(\epsilon)$ and $B \sim {\cal O}(\epsilon^3)$,
with $\epsilon <<1$.
Then, $d B / d T \sim {\cal O}(\epsilon^5) \ll  
-(\beta+\alpha/\beta)B + c A^3$,  
leading to the equation for the center manifold
$B_m(A)= ( - \alpha A/\beta + c A^3) / (\beta + \alpha/\beta) 
+ {\cal O}(\epsilon^5)$. Substituting
this result into Eq. (\ref{eq:vdp5}) gives the normal form equation for 
$A$.

We now turn to the case $\kappa > 0$, and keep the same change of 
variables under the assumption that the intensity of the noise is small:
$\kappa \sim {\cal O}(\epsilon^2)$.
The exact Fokker-Planck equation associated with Eq. (\ref{eq:vdp5}) is,
\begin{eqnarray}
\label{eq:vdp6}
\nonumber
\partial_t  {\cal P}(A,B;t) &=& - \frac{\partial}{\partial A} \left\{ 
\left[
\frac{\alpha}{\beta} (A + B) - c A^3 + d A^2 B + e A B^2 + f B^3 \right] 
{\cal P}(A,B;t)\right\} \\ \nonumber & &
 - \frac{\partial}{\partial B} \left\{ \left[  - \frac{\alpha}{\beta} A - 
\left( \beta + \frac{\alpha}
{\beta} \right) B  + c A^3 - d A^2 B
  - e A B^2 - f B^3 \right] {\cal P}(A,B;t)\right\} \\ & &
+ \left( \frac{\partial^2}{\partial A^2} + \frac{\partial^2}
{\partial B^2} - 2 \frac{\partial^2}{\partial A \partial B} \right)
\left[  \frac{\kappa}{\beta^2} ( A + B)^2 {\cal P}(A,B;t) \right].
\end{eqnarray}
The first step in our analysis is to introduce 
scaled variables $\bar{\kappa} = \kappa/\epsilon^2,\bar A = A /\epsilon^i, 
\bar B = B / \epsilon^j$ and  $\bar{\alpha} = \alpha/\epsilon^{2i}$
in Eq. (\ref{eq:vdp6}).
We choose $\alpha \sim A^2$
in order to have $\alpha A \sim A^3$ in the equation for $A$.
In view of the deterministic result, we further assume that $B/A \ll 1$, 
and thus consider $i<j$. Eq. (\ref{eq:vdp6}) now reads 
\begin{eqnarray}
\label{eq:vdp7}
\nonumber
\partial_t  {{\cal P}(\bar{A},\bar{B};t)} &=& - 
 \frac{\partial}{\partial \bar{A}}
 \left\{ \left[ \epsilon^{2i} \frac{\bar{\alpha}}{\beta} (\bar{A} + 
\epsilon^{j-i} \bar{B}) - \epsilon^{2i} c \bar{A}^3 + \epsilon^{i+j} 
 d \bar{A}^2 \bar{B} + \epsilon^{2j} e \bar{A} \bar{B}^2 + \epsilon^{3j-i} f 
\bar{B}^3 \right] \right. \\ \nonumber & &  \left.
{\cal P}(\bar{A},\bar{B};t) \right\} - \frac{\partial}{\partial \bar{B}} 
 \left\{ \left[  - \epsilon^{2+i-j} \frac{\bar{\alpha}}{\beta} \bar{A} - 
\left( \beta + \frac{\alpha}{\beta} \right) \bar{B}  + 
 \epsilon^{3i-j} c \bar{A}^3 \right. \right. \\ \nonumber & & \left. \left.
 - \epsilon^{2i} d \bar{A}^2 \bar{B}
  - \epsilon^{i+j} e \bar{A} \bar{B}^2 - \epsilon^{2j} f \bar{B}^3 \right] 
 {\cal P}(\bar{A},\bar{B};t) \right\} \\  & &
+ \left( \epsilon^{2-2i} \frac{\partial^2}{\partial \bar{A}^2} 
+ \epsilon^{2-2j} \frac{\partial^2}{\partial \bar{B}^2} 
- 2 \epsilon^{2-(i+j)} 
\frac{\partial^2}{\partial \bar{A} \partial \bar{B}} \right)
\left[ \frac{\bar{\kappa}}{\beta^2} 
(\epsilon^i \bar{A} + \epsilon^j \bar{B})^2 
{\cal P}(\bar{A},\bar{B};t) \right].
\end{eqnarray}
Next, we introduce the decomposition 
${\cal P}(\bar{A},\bar{B};t) = p(\bar{B}|\bar{A};t) P(\bar{A};t)$ 
in Eq. (\ref{eq:vdp7}) and integrate over $\bar{B}$. 
Since $[p(\bar{B}|\bar{A};t)]_{\bar{B}
\rightarrow \pm \infty} =
[\partial_{\bar{B}} p(\bar{B}|\bar{A};t)]_{\bar{B} \rightarrow \pm \infty} 
= 0$, all terms involving a derivative with respect to $\bar{B}$ integrate to
zero, leaving
\be
\label{eq:vdp8}
\partial_t P(\bar{A};t) = - \epsilon^{2i} \frac{\partial}{\partial \bar{A}}
 \left\{ \left(\frac{\bar{\alpha}}{\beta}\bar{A} 
-  c \bar{A}^3 \right) P(\bar{A};t) \right\} +
\epsilon^{2} \frac{\partial^2}{\partial \bar{A}^2} \left\{\frac{\bar{\kappa}}
{\beta^2} \bar{A}^2 P(\bar{A};t) \right\}.
\ee
Only the dominant  contributions to the two terms on the right-hand side 
of Eq. (\ref{eq:vdp8}) were included. In order to obtain a dominant balance at
${\cal O}(\epsilon^2)$, we let $i=1$. The marginal probability 
density $P(\bar{A};t)$ then evolves over a time
scale $T=\epsilon^{2} t$. By contrast, the conditional density 
$p(\bar{B}|\bar{A};t)$ varies over times of ${\cal O}(1)$, as seen from the
equation  
\be
\label{eq:vdp9}
\partial_t p(\bar{B}|\bar{A};t) = - \frac{\partial}{\partial \bar{B}} \left\{
  -\left(\beta + \frac{\alpha}{\beta}\right) \bar{B}
 p(\bar{B}|\bar{A};t) \right\} + \frac{\partial^2}{\partial \bar{B}^2} \left\{
\frac{\bar{\kappa}}{\beta^2} \bar{A}^2 p(\bar{B}|\bar{A};t) \right\},
\ee
obtained by choosing $j=2$ and restricting Eq. ({\ref{eq:vdp7}) to 
${\cal O}(1)$.
The separation of time scales central to the elimination procedure
in deterministic systems is thus preserved in the stochastic case.
The Langevin equation corresponding to Eq. (\ref{eq:vdp9}) is obtained
by setting $A$ to a constant in the original equation for $B$ and dropping
any term of  ${\cal O}(\epsilon^3)$ or higher. 

The stationary solution
to Eq. (\ref{eq:vdp9}) reads, in the original set of variables 
\begin{eqnarray}
\label{eq:vdp10}
p(B|A)= \sqrt{\frac{\beta^2(\alpha/\beta+ \beta)}{2 \pi \kappa A^2}}
\exp \left[ - \frac{\beta^2(\alpha/\beta+ \beta)}{2 \kappa A^2} B^2 \right].
\end{eqnarray}
It is a
Gaussian distribution with zero mean and variance $\sigma^2 (A) =
\kappa A^2 / (\beta + \alpha / \beta) \beta^2$.
The fast variable $B$ thus fluctuates around $B=0$ and not around
the center manifold $B_m(A)$ (in contrast with the results of refs. 
\cite{re:knobloch83} and \cite{re:elphick87}).   
In fact, $B_m(A)/B \sim \epsilon^3/ \epsilon^2 \ll 1$, thus indicating that
terms proportional to $\alpha A$ and $A^3$ in the equation for $B$ 
do not have any significant influence on its evolution.

We note that the statistics of the fast variable are not generic but depend
on the details of the system under consideration. For instance, 
if the equation for the fast variable is 
deterministic, the conditional density is
a delta function on the center manifold \cite{re:drolet97}.
The procedure is then equivalent to replacing $B$ 
in the equation for $A$ by its value on the center manifold. 
Eq. (\ref{eq:vdp10}) also fails if a term proportional to $A^2$ is
present in the equation for $B$, in which case the  
Gaussian distribution is centered on the manifold $B'_m(A) 
\approx \mbox{Const.} \times A^2 \sim {\cal O}(\epsilon^2)$.

The statistical properties of the critical variable $A$ follow 
from Eq. (\ref{eq:vdp8}). In particular,
the stationary solution $P(A)$ (or, equivalently, $P(\bar{A})$) 
to Eq. (\ref{eq:vdp8}) reads
\be
\label{eq:vdp13}
P(A)= {\cal N} |A|^{\frac{\alpha \beta}{\kappa} -2}
    \exp(-c \beta^2 A^2 /2 \kappa).
\ee
This density has nonzero moments and
is normalizable (with ${\cal N} =[c \beta^2/2 \kappa]^{\frac
{\alpha \beta}{2 \kappa} - \frac{1}{2}} /  \Gamma(\frac{\alpha \beta}
{2 \kappa}- \frac{1}{2})$) as long as $\frac{\alpha \beta}{\kappa} -2
> -1$. This implies that, to ${\cal O}(\epsilon^2)$, the bifurcation occurs at
\be
\label{eq:vdp14}
\alpha_c= \kappa / \beta.
\ee
The bifurcation point is thus shifted to positive values of the reduced
control parameter by an amount proportional to the noise intensity $\kappa$.
This result agrees with that of L\"ucke \cite{re:lucke89}
who used a perturbation analysis of the linear stability problem, but
disagrees with earlier results due to Knobloch and Wiesenfeld
\cite{re:knobloch83} and Seshadri, West and Lindenberg \cite{re:seshadri81}.

We next compare our results (Eqs. (\ref{eq:vdp10}) and 
(\ref{eq:vdp13})) with a numerical integration of the original model
equation (Eq. (\ref{eq:vdp4})).
The numerical calculations were performed by using an explicit integration 
scheme, valid
to first order in $\Delta t$ \cite{re:sancho82}, a
step $\Delta t =0.005$ and a bin size for the various probability
densities $\Delta A =0.01$ and $\Delta B= 0.001$. 
Initial conditions for $x$ and 
$\dot{x}$ were chosen randomly
from a uniform distribution in the interval $[-0.5,0.5]$. Results
from 100 independent runs were averaged, and within each run the 
various quantities were sampled every 1000 steps.
To ensure the system had reached a stationary state, the first one million
steps were discarded.
Just above onset, the density Eq. (\ref{eq:vdp13}) exhibits a divergence
at the origin (Fig. \ref{fig:f1}A). At $\alpha = 2 \kappa / \beta$,
this divergence
transforms into a maximum (Fig. \ref{fig:f1}B) which moves to higher values
of $A$ as the control parameter
is further increased (Fig. \ref{fig:f1}C). All three figures,
corresponding to
the parameter values $\beta=a=b=1$ and $\kappa=0.01$, show excellent
agreement between the predictions of
Eq. (\ref{eq:vdp13}) and the stationary densities computed numerically.

The average amplitude $\langle |A| \rangle$ was also computed for various
values of $\alpha$, and the results compared with
the analytic result
\be
\langle |A| \rangle = \left[ \frac{c \beta^2}{2 \kappa} \right]^{-1/2}
 \frac{\Gamma(\alpha \beta/ 2 \kappa)}{\Gamma(\alpha \beta/ 2 \kappa - 1/2)},
\ee
which follows from Eq. (\ref{eq:vdp13}). As shown in
Fig. \ref{fig:f1}D, agreement between the two data sets
is once again excellent.
As an additional test, we considered the statistics for the fast variable $B$.
Combining Eqs. (\ref{eq:vdp10}) and (\ref{eq:vdp13}) we  find
\be
\langle |B| \rangle = 2 \int_{-\infty}^\infty P(A) dA \int_0^\infty B
  p(B|A) dB = \sqrt{\frac{ 2 \kappa}{\pi \beta^2
 (\alpha/\beta+ \beta)}} \langle |A| \rangle.
\ee
Analytic and simulation results are compared in Fig. \ref{fig:f2}.
The quantity $P(|B|)$ plotted in Figs. \ref{fig:f2}A, \ref{fig:f2}B and
\ref{fig:f2}C  is the density of probability of
finding some value of $|B|$ independently of the value of $A$, i.e.,
$P(|B|) = 2 P(B) = 4 \int_0^\infty P(A)p(B|A) dA$.
As before, both sets of results agree extremely well.

\section{Transcritical bifurcation with multiplicative noise}
\label{sec:transc}

As a second illustration of the approach, we study 
the set of two equations
\begin{eqnarray}
\label{eq:tr1}
    \frac{d}{dt} \left[ \begin{array}{c} A \\  B
        \end{array}  \right]  & = & \left[ \begin{array}{cc}
         \alpha  &  0  \\
          0  & -\lambda \end{array} \right]
        \left[ \begin{array}{c} A \\ B \end{array} \right] +
        \left[ \begin{array}{c}
           -a A^2  - b A B - c B^2  \\
           +d A^2  + e A B + f B^2  \end{array} \right]
      + \left[ \begin{array}{cc} m_{11} & m_{12} \\ 
        m_{21} & m_{22}  \end{array} \right] \left[ 
      \begin{array}{c} A \\ B \end{array} \right]
         \xi(t),
\end{eqnarray}
with $\alpha$ small and all the remaining coefficients of ${\cal O}(1)$. 
In the deterministic limit, the variable $B$ relaxes quickly to the 
center manifold $B_m(A)=d A^2 /\lambda$, and the normal form equation
is given by 
\be
\label{eq:tr2}
\frac{dA}{dt} = \alpha A - a A^2,
\ee
which describes a transcritical bifurcation at $\alpha=0$.
Following the procedure introduced above,
we define the rescaled parameters $\bar{\kappa} = \kappa/\epsilon^2$ 
and $\bar{\alpha} = \alpha/\epsilon^i$, and rescaled variables 
$\bar{A}=A/\epsilon^i$ and $\bar{B}=B/\epsilon^j$, with $i<j$.  
The exact Fokker-Planck equation corresponding to 
Eq. (\ref{eq:tr1}) then reads
\begin{eqnarray}
\label{eq:tr3}
\nonumber
 \partial_t {\cal P}(\bar{A},\bar{B};t) 
 & = & - \frac{\partial}{\partial \bar{A}} \{ [
 \epsilon^{i} \bar{\alpha} \bar{A}  - \epsilon^{i} a \bar{A}^2 
- \epsilon^{j} b \bar{A} \bar{B} - 
\epsilon^{2j-i} c \bar{B}^2 + \epsilon^{2} \bar{\kappa} m_{11} (
 m_{11} \bar{A} +  \epsilon^{j-i} m_{12} \bar{B})   
 \\ \nonumber & & + \epsilon^{2} \bar{\kappa} m_{12} (m_{21} \bar{A} +  
\epsilon^{j-i} m_{22} \bar{B}) ] {\cal P}
(\bar{A},\bar{B})\} 
 - \frac{\partial}{\partial \bar{B}} \{ [ - \lambda \bar{B} + \epsilon^{2i-j} 
 d \bar{A}^2 +  \epsilon^{i} e \bar{A} \bar{B}  +  \epsilon^{j} f \bar{B}^2 
 \\ \nonumber & &  + \epsilon^{2} \bar{\kappa} m_{22} ( \epsilon^{i-j} m_{21} 
 \bar{A} +  m_{22} \bar{B})  + \epsilon^{2} \bar{\kappa} m_{21} 
(\epsilon^{i-j} m_{11} \bar{A} 
+ m_{12} \bar{B}) ] 
 {\cal P}(\bar{A},\bar{B})\} \\ \nonumber & &
+  \frac{\partial^2}{\partial \bar{A}^2} \{ [ \epsilon^{2-2i} \bar{\kappa}
(\epsilon^{i} m_{11} \bar{A} + \epsilon^{j} m_{12} \bar{B})^2 ] 
{\cal P}(\bar{A},\bar{B})  \} \\ \nonumber & & 
+ \frac{\partial^2}{\partial \bar{B}^2} \{ [ \epsilon^{2-2j} \bar{\kappa}
(\epsilon^{i} m_{21} \bar{A} +  \epsilon^{j} m_{22} \bar{B})^2 ] 
{\cal P}(\bar{A},\bar{B})  \} \\  & &
+ 2  \frac{\partial^2}{\partial \bar{A} \partial \bar{B}}
 \{[ \epsilon^{2-i-j} \bar{\kappa} (\epsilon^{i} m_{11} \bar{A} +  
\epsilon^{j} m_{12} \bar{B}) 
 (\epsilon^{i} m_{21} \bar{A} + \epsilon^{j} m_{22} \bar{B}) ] 
{\cal P}(\bar{A},\bar{B}) \}. 
\end{eqnarray}
Integrating this equation over $\bar{B}$ gives
\begin{eqnarray}
\label{eq:tr3b}
\nonumber \partial_t  P(\bar{A};t) & = &
  -  \frac{\partial}{\partial \bar{A}} \left\{ [
  \epsilon^i \bar{\alpha} \bar{A}  - \epsilon^i a \bar{A}^2 + 
   \epsilon^{2} \bar{\kappa} (m_{11}^2 + m_{12} m_{21}) \bar{A} 
  ] P(\bar{A};t)\right\} \\ & &  
  + \frac{\partial^2}{\partial \bar{A}^2} \left\{  \epsilon^2 \bar{\kappa}
   m_{11}^2 \bar{A}^2  P(\bar{A};t)  \right\}. 
\end{eqnarray}
As in Section \ref{sec:vdp}, only the leading contributions to the 
right-hand side of Eq. (\ref{eq:tr3b}) were included. In order to 
have a dominant balance at ${\cal O}(\epsilon^2)$, we choose $i=2$. 
Similarly, letting $j=3$ in Eq. (\ref{eq:tr3}) leads to the equation 
\be
\label{eq:tr4}
\partial_t  p(\bar{B}|\bar{A};t) = - \frac{\partial}{\partial \bar{B}}
[ - \lambda \bar{B} p(\bar{B}|\bar{A})] + \frac{\partial^2}
{\partial \bar{B}^2} [ \bar{\kappa} m_{21}^2 \bar{A}^2 p(\bar{B}|\bar{A})],
\ee
valid to ${\cal O}(1)$.
Eqs. (\ref{eq:tr3b}) and (\ref{eq:tr4}) admit the stationary solutions  
\be
\label{eq:tr5}
P(\bar{A}) =  {\cal N} \bar{A}^{\frac{\bar{\alpha}}{\bar{\kappa} m_{11}^2}
 + \frac{m_{12} m_{21}}
{m_{11}^2} -1} \exp \left( - \frac{a}{\bar{\kappa} m_{11}^2} \bar{A} \right),
\ee
with ${\cal N} = (a/\bar{\kappa} m_{11}^2)^{\frac{\bar{\alpha}}{\bar{\kappa}
 m_{11}^2} +\frac{m_{12} m_{21}}{m_{11}^2}+1} / 
\Gamma(\frac{\bar{\alpha}}{\bar{\kappa} m_{11}^2}
+\frac{m_{12} m_{21}}{m_{11}^2}+1)$, and
\be
\label{eq:tr5b}
p(\bar{B}|\bar{A}) = \sqrt{\frac{\lambda}{2 \pi \bar{\kappa} m_{21}^2
 \bar{A}^2}} \exp \left[ - \frac{\lambda \bar{B}^2}{2
\bar{\kappa} m_{21}^2 \bar{A}^2} \right]
\ee
respectively. The normalizability condition 
$\bar{\alpha}/\bar{\kappa} m_{11}^2 +m_{12} m_{21}/
{m_{11}^2}-1 > -1$ associated with Eq. (\ref{eq:tr5}) places the 
bifurcation point at 
\be
\label{eq:tr5c}
\alpha_c = -\kappa m_{12} m_{21},
\ee
valid to ${\cal O}(\epsilon^2)$.
Predictions from Eq. (\ref{eq:tr5}) are compared with numerical estimates
obtained through direct integration of Eq. (\ref{eq:tr1}) in
Fig. \ref{fig:f3}.
The computations were performed with $\kappa=0.01$ and
all the parameters in the original equations except $\alpha$ set to one.
The numerical and analytical estimates 
(represented by black dots and  solid lines respectively) 
are virtually indistinguishable near onset. Significant differences
do appear, however, as $\alpha$ increases.  
Results pertaining to the fast variable $B$ are
presented in Fig. \ref{fig:f4}. Again, analytic estimates of
$P(B) =  \int_{0}^{+\infty} P(A) p(B|A) dA$ obtained by using
Eqs. (\ref{eq:tr5}) and (\ref{eq:tr5b}) compare well
with their numerical counterparts near onset, but become
increasingly inaccurate as $\alpha$ increases. 
In particular, the numerical results indicate that
the density $P(B)$ is slightly skewed and has a non-zero
average. These properties are incompatible with
a distribution such as Eq. (\ref{eq:tr5b}) which is even in $B$.

We show next that it is in principle straightforward to systematically
improve the accuracy of the analytic calculation by going to higher 
orders in $\epsilon$. We seek a stationary solution of Eq. (\ref{eq:tr3}) 
valid to one more order in $\epsilon$ (${\cal O}(\epsilon)$). 
Setting $\partial_t {\cal P}(A,B;t) =0$, $i=2$, and $j=3$  
in this equation and keeping terms up
to ${\cal O}(\epsilon)$ gives, after some algebra,   
\begin{eqnarray}
\label{eq:tr6}
\nonumber 0 & = &  -[ - \lambda \bar{B} + \epsilon 
 d \bar{A}^2 - \epsilon \bar{\kappa} (m_{22} m_{21} + 3 m_{21} m_{11}) 
  \bar{A}]  P(\bar{A}) p(\bar{B}|\bar{A}) \\ \nonumber & &
+ [\bar{\kappa} m_{21}^2 \bar{A}^2 + \epsilon (2 \bar{\kappa} m_{21} m_{22} 
 \bar{A} \bar{B})] P(\bar{A}) \frac{\partial p(\bar{B}|\bar{A})}
  {\partial \bar{B}} + \epsilon (2 \bar{\kappa} m_{11} m_{21} \bar{A}^2)
 \\ & &  \left[ p(\bar{B}|\bar{A}) 
 \frac{d P(\bar{A})}{d \bar{A}}
 +P(\bar{A}) \frac{\partial p(\bar{B}|\bar{A})}{\partial \bar{A}} \right].
\end{eqnarray}
In contrast with the calculation above, the equations for the 
conditional and marginal probabilities do not decouple.
In order to solve Eq. (\ref{eq:tr6}) for $p(\bar{B}|\bar{A})$, 
the derivatives $d P(\bar{A})/ d\bar{A}$ and $\partial 
p(\bar{B}|\bar{A})/\partial \bar{A}$ must be known to ${\cal O}(1)$.

We first 
determine $dP(\bar{A})/d\bar{A}$ by noting that the stationary solution
to Eq. (\ref{eq:tr3b}) satisfies 
\be
\label{eq:tr7}
\bar{\kappa} m_{11}^2 \bar{A}^2 \frac{d P(\bar{A})}{d \bar{A}} - \left[
(\bar{\alpha} - \bar{\kappa} m_{11}^2 + \bar{\kappa} m_{12} m_{21}) \bar{A}
-a \bar{A^2} \right] P(\bar{A}) = 0.
\ee
We also assume that the conditional
probability density $p(\bar{B}|\bar{A})$ is of the form
\be
\label{eq:tr7b}
p(\bar{B}|\bar{A}) = {\cal N}_\epsilon \exp \left[ - \frac{\lambda \bar{B}^2}{2
\bar{\kappa} m_{21}^2 \bar{A}^2} + {\cal O}(\epsilon) \right],
\ee
i.e., that the improved calculation simply adds  corrections
to the argument of the exponential in Eq. (\ref{eq:tr5b}). 
Under that assumption, 
\be
\label{eq:tr8}
\frac{\partial p(\bar{B}|\bar{A})}{\partial \bar{A}} = - \frac{p(\bar{B}|\bar{A}
)}
{\bar{A}} + \frac{\lambda \bar{B}^2}{\bar{\kappa} m_{21}^2 \bar{A}^3}
p(\bar{B}|\bar{A}) + {\cal O}(\epsilon).
\ee
Combining Eqs. (\ref{eq:tr6}), (\ref{eq:tr7}) and (\ref{eq:tr8}) gives,  
to ${\cal O}(\epsilon)$,
\begin{eqnarray}
\label{eq:tr9}
\nonumber 0 & = & 
[\bar{\kappa} m_{21}^2 \bar{A}^2 + \epsilon (2 \bar{\kappa} m_{21} m_{22}
 \bar{A} \bar{B})] \frac{\partial p(\bar{B}|\bar{A})}{\partial \bar{B}}
- \left\{ -\lambda \bar{B} + \epsilon \left( d + \frac{2 a m_{21}}{m_{11}} 
\right) \bar{A}^2   \right. \\  & & \left.- \epsilon
\left[ \frac{2 m_{21}}{m_{11}}(\bar{\alpha} + \bar{\kappa}
m_{12} m_{21}) 
 + \frac{}{} \bar{\kappa} m_{22} m_{21}  - 
\bar{\kappa} m_{21} m_{11}
\right] \bar{A} - \epsilon \frac{2 m_{11} \lambda \bar{B}^2}{m_{21} \bar{A}}
 \right\} p(\bar{B}|\bar{A}).
\end{eqnarray}
The solution to that equation is an exponential, the argument of which 
can be expanded in the small quantity $\bar{B}/\bar{A}$. This
yields the probability density 
\begin{eqnarray}
\label{eq:tr10}
\nonumber p(\bar{B}|\bar{A}) & = & {\cal N}_\epsilon \exp \left\{ 
- \frac{\lambda}{2 \bar{\kappa} m_{21}^2} \left( \frac{\bar{B}}{\bar{A}} 
\right)^2 + \epsilon \frac{2 (m_{22}-m_{11}) \lambda}{3 \bar{\kappa} m_{21}^3}
\left( \frac{\bar{B}}{\bar{A}} \right)^3 + \epsilon   
\left[  \left( d + \frac{2 a m_{21}}{m_{11}} \right) \frac{\bar{A}}
{\bar{\kappa} m_{21}^2} \right.  \right. \\ 
 & & \left. \left. -
  \frac{2 m_{21}}{m_{11}}(\bar{\alpha} + \bar{\kappa} m_{12} m_{21}) - 
 \bar{\kappa} m_{22} m_{21}   + \bar{\kappa} m_{21} m_{11} \right] 
\frac{\bar{B}}{\bar{A}} \right\},
\end{eqnarray}
which is consistent with the assumption in Eq. (\ref{eq:tr7b}).

Note that the presence of a cubic term in the exponential  
implies that the fast variable exhibits 
non Gaussian fluctuations. A divergence at either $\bar{B} \rightarrow - 
\infty$ or $\bar{B} \rightarrow + \infty$ also means 
that Eq. (\ref{eq:tr10}) is non-normalizable.
In practice, however, one can compute an effective normalization
constant by integrating Eq. (\ref{eq:tr10}) over some interval 
$[\bar{B}_-,\bar{B}_+]$ 
at the limits of which $p(\bar{B}_\pm|\bar{A}) \ll 1$. 
Alternatively, higher order terms could be included in the Taylor series 
expansion leading to Eq. (\ref{eq:tr10}).
For simplicity however, we let 
$m_{22}=m_{11}$, in which case the coefficient of the cubic term
vanishes and Eq. (\ref{eq:tr10}) simplifies to
\begin{eqnarray}
\label{eq:tr11}
\nonumber 
 p(\bar{B}|\bar{A}) & = &  \sqrt{\frac{\lambda}{2 \pi \bar{\kappa} m_{21}^2
 \bar{A}^2}}  \exp \left\{ - \frac{\lambda}{2 \bar{\kappa} m_{21}^2 \bar{A}^2}
\left[ \bar{B} -  \epsilon \left( d + \frac{2 a m_{21}}{m_{11}} \right) 
\frac{\bar{A}^2}{\lambda} \right. \right.  \\ & & \left. \left. 
- \epsilon \frac{2 m_{21}}{\lambda m_{11}}(\bar{\alpha} + \bar{\kappa} 
m_{12} m_{21}) \bar{A} \right]^2 \right\}.
\end{eqnarray}
Eq. (\ref{eq:tr11}) is a Gaussian distribution with  mean
$\langle \bar{B} \rangle_{\bar{A}} = \int_{-\infty}^{+\infty} \bar{B} 
p(\bar{B}|\bar{A}) d\bar{B} = \epsilon 
\left( d + \frac{2 a m_{21}}{m_{11}} \right) \frac{\bar{A}^2}{\lambda}
 + \epsilon \frac{2 m_{21}}{\lambda m_{11}}(\bar{\alpha} + 
 \bar{\kappa} m_{12} m_{21})
 \bar{A}$ different from  the center manifold $\bar{B}_m(\bar{A})$,
and variance $\langle \bar{B}^2 \rangle_{\bar{A}} = \bar{\kappa} m_{21}^2
\bar{A}^2 /\lambda$. 

As before, we determine the stationary properties of the slow variable 
$\bar{A}$ by setting
$\partial_t {\cal P}(\bar{A},\bar{B};t)=0$ in Eq. (\ref{eq:tr3})  
and integrating 
over $\bar{B}$. The resulting equation reads
\begin{eqnarray}
\label{eq:tr12}
\nonumber 0 & = & \frac{d}{d \bar{A}} \left\{ \left[ \bar{\kappa} m_{11}^2 
\bar{A}^2 + \epsilon (2 \bar{\kappa} m_{11} m_{12} \bar{A} 
\langle \bar{B} \rangle_{\bar{A}})
+ \epsilon^2 \bar{\kappa} m_{22}^2  \langle \bar{B}^2 \rangle_{\bar{A}}
 \right] P(\bar{A}) \right\} \\ 
& &  - \left[
(\bar{\alpha} + \bar{\kappa} m_{11}^2 + \bar{\kappa} m_{12} m_{21}) \bar{A} 
-a \bar{A}^2 + \epsilon (2 \kappa m_{11} m_{12} -b \bar{A}) 
 \langle \bar{B} \rangle_{\bar{A}} - \epsilon^2 c \langle \bar{B}^2 
\rangle_{\bar{A}} \right] P(\bar{A}).
\end{eqnarray}
Inserting the expressions derived above for  $\langle \bar{B} 
\rangle_{\bar{A}}$ and  $\langle \bar{B}^2 \rangle_{\bar{A}}$
in that equation, and rearranging the various terms, we find
\begin{eqnarray}
\label{eq:tr13}
\nonumber 0 & = & -\{[\bar{\alpha} - \bar{\kappa} m_{11}^2 + 
\bar{\kappa} m_{12} m_{21}
 - \epsilon^2 (2 \bar{\kappa} m_{11} m_{21} q) ] \bar{A} 
 - [a + \epsilon^2 (b q + 4 \bar{\kappa} m_{11} m_{12} r + s)] \bar{A}^2 
  \\ & & - \epsilon^2 b r \bar{A}^3 \} P(\bar{A}) + \bar{A}^2 
  [\bar{\kappa} m_{11}^2 + \epsilon^2 ( 2 \bar{\kappa} m_{11} m_{12} r \bar{A} 
  + 2 \bar{\kappa} m_{11} m_{21} q +  \bar{\kappa}^2 m_{12}^2 m_{21}^2
 /\lambda)] \frac{d P(\bar{A})}{d \bar{A}},
\end{eqnarray}
with $q=-2 m_{21} (\bar{\alpha}+ \bar{\kappa} m_{12} m_{21})/\lambda m_{11},
r=(d+2 a m_{21}/m_{11})/\lambda$ and $s=c \bar{\kappa} m_{21}^2/\lambda$.
An approximate solution can be found that, in the original variables,
reads
\be
\label{eq:tr15}
P(A)  =  {\cal N}^{\prime}_\epsilon A^{\mu}  \exp 
\left( \frac{u r}{\kappa m_{11}^2} A^2
 - \frac{v}{\kappa m_{11}^2} A \right),
\ee 
with
\begin{equation}
\mu = \frac{\alpha + \kappa m_{21} m_{12} - \kappa m_{11}^2 
- 2 \kappa m_{11} m_{12} q' -2 \kappa^2 m_{12}^2 m_{21}^2 /\lambda}
{\kappa m_{11}^2 + 2 \kappa m_{11} m_{12} q'
+ \kappa^2 m_{12}^2 m_{21}^2/\lambda},
\end{equation}
$u = a m_{12}/m_{11} - b/2$, 
$$
v = a + \kappa c m_{21}^2/\lambda + 
(b - 2 a m_{12}/m_{11}) q' + 2(\alpha +\kappa
m_{12} m_{21} + \kappa m_{11}^2) m_{12} r/m_{11} - 
\kappa a m_{12}^2 m_{21}^2/ \lambda m_{11}^2
$$ 
and $q' = 
-2 m_{21} (\alpha+ \kappa m_{12} m_{21})/\lambda m_{11}$. 
The dashed lines in Figs. \ref{fig:f3} and \ref{fig:f4} 
are analytic estimates computed
using  Eqs. (\ref{eq:tr11}) and (\ref{eq:tr15}). As expected, 
comparison with the numerical results 
shows a net improvement from our previous predictions (Eqs. (\ref{eq:tr5}) and 
(\ref{eq:tr5b})).

We conclude our analysis by discussing briefly the mechanism by which the
effective bifurcation point differs from its deterministic location in the 
two cases studied above. Consider first the Van der Pol-Duffing oscillator,
Eq. (\ref{eq:vdp5}), which we rewrite as 
\begin{eqnarray}
\label{eq:bla1}
    \frac{d}{dt} \left[ \begin{array}{c} A \\  B
        \end{array}  \right]  & = & \left[ \begin{array}{cc}
         \alpha'  &  0  \\
           0 & - \beta + \alpha' \end{array} \right]
        \left[ \begin{array}{c} A \\ B \end{array} \right] +
        \left[ \begin{array}{c}
           -c A^3   \\ 0 \end{array} \right]
      + \left[ \begin{array}{cc} m_{11} & m_{12} \\ m_{21} & m_{22} 
     \end{array} \right]
        \left[ \begin{array}{c} A \\ B \end{array} \right] \xi(t),
\end{eqnarray}
with $\alpha'= \alpha/\beta, m_{11}=m_{12}=1/\beta$ and 
$m_{21}=m_{22}=-1/\beta$. For simplicity, we only include
in the deterministic part of Eq. (\ref{eq:bla1}) the terms 
which were found relevant in Section \ref{sec:vdp}. 
Note that since $A$ varies over a time scale $T=\epsilon^2 t$, the term
proportional to $B \xi$ in  the equation governing
its evolution can be averaged over the fast time scale. 
With the introduction of the 
scaled variables $A=\epsilon \bar{A}, B =\epsilon^2 \bar{B}, 
{\alpha'}=\epsilon^2 \bar{\alpha}', \kappa= \epsilon^2 \bar{\kappa}$ 
and $T=\epsilon^2 t$, 
this gives 
\be
\label{eq:bla2}
\epsilon^3 \frac{d \bar{A}}{d T} = \epsilon^3 \bar{\alpha}' \bar{A} - 
\epsilon^3 \bar{A}^3 + \epsilon^2 m_{12} \langle \bar{B} \xi \rangle + 
\epsilon m_{11} \bar{A} \xi (T),
\ee
where we have approximated the temporal average of $\bar{B} \xi$ by its
ensemble average, and
where $\langle \xi (T) \xi (T') \rangle = 2 \epsilon^4 \bar{\kappa} 
\delta (T - T')$. 
By using the Furutsu-Novikov theorem \cite{re:furutsu63,re:novikov65},
we find
\be
\label{eq:bla3}
\langle \bar{B} \xi \rangle = \langle \bar{B} \rangle \langle \xi \rangle
+ \epsilon^2 \bar{\kappa} \left\langle \frac{\delta \bar{B}}{\delta \xi} 
 \right\rangle = \epsilon \bar{\kappa}  m_{21} \bar{A} + \ldots
\ee
Therefore, the correlation of $B \xi$ itself evolves
over the slow time scale. 
By combining Eqs. (\ref{eq:bla2}) and (\ref{eq:bla3}) we obtain
the effective normal form equation,
\be
\label{eq:bla4}
 \frac{d \bar{A}}{d T} = \tilde{\alpha} \bar{A} 
 - \bar{A}^3 + \epsilon^{-2} m_{11} \bar{A} \xi (T),
\ee
with $\tilde{\alpha} = \bar{\alpha}' + \bar{\kappa}  m_{21} m_{12}$.
This equation also leads to the Fokker-Planck equation
(Eq. (\ref{eq:vdp8})) already derived in Section \ref{sec:vdp}.
The bifurcation point associated with Eq. (\ref{eq:bla4}) is located at
$\tilde{\alpha}=0$ \cite{re:schenzle79,re:graham82b}, 
i.e., at  
\be
\label{eq:bla5}
\alpha' = -\kappa  m_{12} m_{21},
\ee
in agreement with our previous result (Eq. (\ref{eq:vdp14})). 
Equation (\ref{eq:bla5}) is also identical to Eq. (\ref{eq:tr5c}) derived
in the case of a generic 
transcritical bifurcation. Hence, to first order in the intensity of the
noise, the location 
of the bifurcation point is entirely determined 
from the stochastic part of the original set of equations and is therefore
independent of the nature of the bifurcation. 
Corrections to Eq. (\ref{eq:bla5}), however, depend on the 
details of the system under consideration.
For instance, in the case of the transcritical
bifurcation, the first correction to Eq. (\ref{eq:bla5}) follows from
the normalizability condition $\mu > -1$ associated with  
the probability density $P(A)$ given in Eq. (\ref{eq:tr15}). 

\section*{Acknowledgments}
This research was supported by the Microgravity Science and Applications
Division of the NASA under contract No. NAG3-1885, and also in part by the 
Supercomputer Computations Research Institute, which is partially funded by 
the U.S.  Department of Energy, contract No. DE-FC05-85ER25000.


\begin{thebibliography}{10}

\bibitem{re:guckenheimer83}
J. Guckenheimer and P. Holmes, {\em Nonlinear oscillations, dynamical systems,
  and bifurcations of vector fields} (Springer Verlag, New York, 1983).

\bibitem{re:stratonovich67}
R. Stratonovich, {\em Topics in the Theory of Random Noise} (Gordon and Breach,
  New York, 1967), Vol.~II.

\bibitem{re:haken77}
H. Haken, {\em Synergetics: an introduction} (Springer, Berlin, 1977).

\bibitem{re:vandenbroeck82}
C. van~den Broeck, M. {Malek~Mansour}, and F. Baras, J. Stat. Phys. {\bf 28},
  557  (1982).

\bibitem{re:knobloch83}
E. Knobloch and K. Wiesenfeld, J. Stat. Phys. {\bf 33},  611  (1983).

\bibitem{re:elphick87}
C. Elphick, M. Jeanneret, and E. Tirapegui, J. Stat. Phys. {\bf 48},  925
  (1987).

\bibitem{re:drolet97}
F. Drolet and J. {Vi\~nals}, Phys. Rev. E {\bf 56},  2649  (1997).

\bibitem{re:schenzle79}
A. Schenzle and H. Brand, Phys. Rev. A {\bf 20},  1628  (1979).

\bibitem{re:graham82b}
R. Graham and A. Schenzle, Phys. Rev. A {\bf 25},  1731  (1982).

\bibitem{re:lucke89}
M. {L\"{u}cke},  in {\em Noise in nonlinear dynamical systems}, Vol.~2 of {\em
  Theory of noise induced processes in special applications}, edited by F. Moss
  and P. McClintock (Cambridge University Press, Cambridge, 1989).

\bibitem{re:seshadri81}
V. Seshadri, B. West, and K. Lindenberg, Physica {\bf 107A},  219  (1981).

\bibitem{re:wiesenfeld82}
K. Wiesenfeld and E. Knobloch, Phys. Rev. A {\bf 26},  2946  (1982).

\bibitem{re:sancho82}
J. Sancho, M.~S. Miguel, S. Katz, and J. Gunton, Phys. Rev. A {\bf 26},  1589
  (1982).

\bibitem{re:furutsu63}
K. Furutsu, J. Res. Nat. Bur. Standards {\bf 67D},  303  (1963).

\bibitem{re:novikov65}
E. Novikov, Sov. Phys. JETP {\bf 20},  1290  (1965).

\end{thebibliography}

\begin{figure}[p]
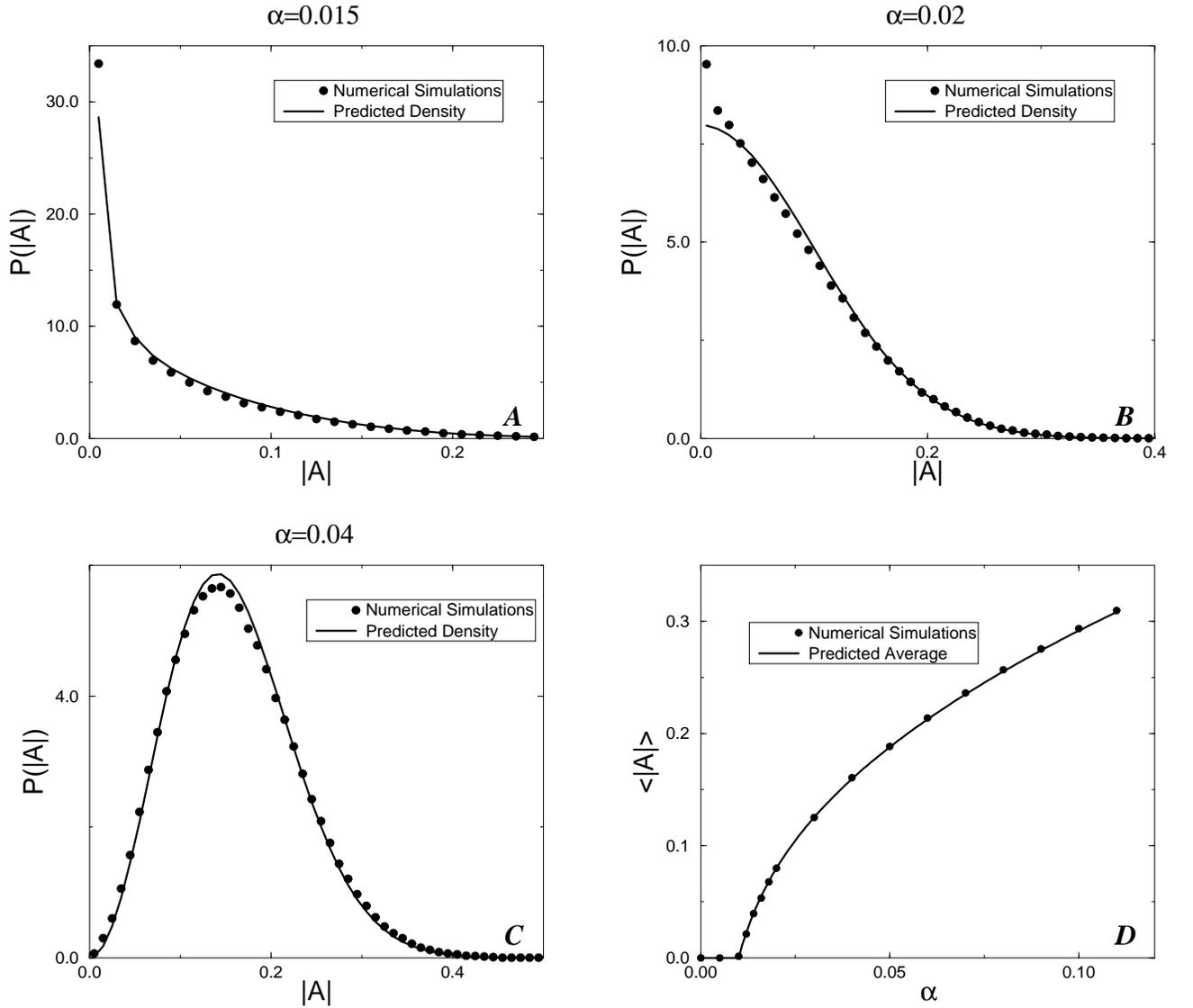

\vbox{ \hbox{
     \psfig{figure=fd2.fig1a,width=3.5in}
     \psfig{figure=fd2.fig1b,width=3.5in}}
       \hbox{
     \psfig{figure=fd2.fig1c,width=3.5in}
     \psfig{figure=fd2.fig1d,width=3.5in}}
      }
\caption{(A), (B) and (C): stationary probability densities as a
function of the absolute value of $A$ for the Van der Pol-Duffing
oscillator. Shown are the densities above onset for three 
different values of the control parameter $\alpha$.
(A), $\alpha=0.015$; (B), $\alpha=0.02$; and, (C), $\alpha=0.04$. We
show in (D) the bifurcation diagram showing the average value $\langle |A| 
\rangle$ as a function of $\alpha$. In all cases, the analytic results
are represented by a solid line, whereas the symbols are the results of the
numerical calculation.}
\label{fig:f1}
\end{figure}

\newpage

\begin{figure}[p]
\vbox{ \hbox{
     \psfig{figure=fd2.fig2a,width=3.5in}
     \psfig{figure=fd2.fig2b,width=3.5in}}
       \hbox{
     \psfig{figure=fd2.fig2c,width=3.5in}
     \psfig{figure=fd2.fig2d,width=3.5in}}
      }
\caption{(A), (B) and (C): stationary probability densities as a
function of the absolute value of $B$ for the Van der Pol-Duffing
oscillator. Shown are the densities above onset for three 
different values of the control parameter $\alpha$.
(A), $\alpha=0.015$; (B), $\alpha=0.02$; and, (C), $\alpha=0.04$. We
show in (D) the bifurcation diagram showing the average value $\langle |B| 
\rangle$ as a function of $\alpha$. In all cases, the analytic results
are represented by a solid line, whereas the symbols are the results of the
numerical calculation.}
\label{fig:f2}
\end{figure}

\newpage 

\begin{figure}[p]
\vbox{ \hbox{
     \psfig{figure=fd2.fig3a,width=3.5in}
     \psfig{figure=fd2.fig3b,width=3.5in}}
       \hbox{
     \psfig{figure=fd2.fig3c,width=3.5in}
     \psfig{figure=fd2.fig3d,width=3.5in}}
      }
\caption{(A), (B) and (C): stationary probability densities as a
function of $A$ for the transcritical bifurcation.
Shown are the densities above onset for three 
different values of the control parameter $\alpha$.
(A), $\alpha=-0.005$; (B), $\alpha=0.01$; and, (C), $\alpha=0.04$. We
show in (D) the bifurcation diagram showing $\langle A 
\rangle$ as a function of $\alpha$. In all cases, the analytic results
to ${\cal O}(1)$ and ${\cal O}(\epsilon^{2})$ are represented by solid
and dashed lines respectively, whereas the symbols are the results of the
numerical calculation.}
\label{fig:f3}
\end{figure}

\newpage

\begin{figure}[p]
\vbox{ \hbox{
     \psfig{figure=fd2.fig4a,width=3.5in}
     \psfig{figure=fd2.fig4b,width=3.5in}}
       \hbox{
     \psfig{figure=fd2.fig4c,width=3.5in}
     \psfig{figure=fd2.fig4d,width=3.5in}}
      }
\caption{(A), (B) and (C): stationary probability densities as a
function of $B$ for the transcritical bifurcation.
Shown are the densities above onset for three 
different values of the control parameter $\alpha$.
(A), $\alpha=-0.005$; (B), $\alpha=0.01$; and, (C), $\alpha=0.04$. We
show in (D) the bifurcation diagram showing $\langle B 
\rangle$ as a function of $\alpha$. In all cases, the analytic results
to ${\cal O}(1)$ and ${\cal O}(\epsilon)$ are represented by solid
and dashed lines respectively, whereas the symbols are the results of the
numerical calculation.}
\label{fig:f4}
\end{figure}

\end{document}